\documentclass[fleqn]{article}
\usepackage{longtable}
\usepackage{graphicx}

\begin {document}
\begin{flushleft}
{\LARGE
{\bf Comment on ``Electron impact excitation and ionization cross section of tungsten ions, W$^{44+}$" by El-Maaref et al.\,   [J. Quant. Spectrosc. Radiat. Transfer 2019, 224:147]}
}\\

\vspace{1.5 cm}

{\bf {Kanti  M  ~Aggarwal}}\\ 

\vspace*{1.0cm}

Astrophysics Research Centre, School of Mathematics and Physics, Queen's University Belfast, \\Belfast BT7 1NN, Northern Ireland, UK\\ 
\vspace*{0.5 cm} 

e-mail: K.Aggarwal@qub.ac.uk \\

\vspace*{0.20cm}

Received: 14 March 2019; Accepted: 12 April 2019

\vspace*{1.0 cm}

\vspace*{1.0 cm}
\vspace*{1.0 cm}

\hrule

\vspace{0.5 cm}

\end{flushleft}

\clearpage


\begin{abstract}

In a recent paper, El-Maaref  et al.  [J. Quant. Spectrosc. Radiat. Transfer 2019, 224:147] have reported  atomic data for Zn-like W~XLV. Their results are mainly for  energy levels, radiative rates, collision strengths  ($\Omega$) for electron impact excitation, and cross sections for ionization, but  for  only for a few levels/transitions, and in a very limited range of energy. For the calculations of $\Omega$ they have adopted the DARC code. In this comment, through our independent calculations, we demonstrate that their results for $\Omega$ are highly underestimated by over two orders of magnitude, and hence are totally unreliable.

\end{abstract}

\clearpage

\section{Introduction}

Tungsten (W) is perhaps the most important element for the construction of fusion reactors and it also radiates at almost all ionization stages. Therefore, the study of its various properties is very important, particularly for the success of the developing ITER (International Thermonuclear Experimental Reactor) project. In this direction, several scientists are making efforts in producing atomic data, theoretically as well as experimentally, for many parameters, such as energy levels, radiative rates and collision strengths  ($\Omega$). Although experimental results  are very useful for the validation of theoretical data, they are often very limited in the energy range and/or the number of levels/transitions. Therefore, to fulfil the requirements data are generated theoretically through the various methods and codes available in the literature. Towards this goal, El-Maaref  et al. \cite{elm} have recently reported atomic data for one of the important ion, namely Zn-like W~XLV. The parameters for which they have reported results are mainly energy levels, radiative rates, collision strengths  ($\Omega$) for electron impact excitation, and cross sections for ionization. However, their results are  for  only  a few levels/transitions, and in a very limited range of energy, which highly restricts the use in any application. More importantly, their  reported data, particularly for $\Omega$, do not appear to be correct, and therefore in this comment we demonstrate the deficiency and unreliability of their results, and also provide reasons for the noted discrepancies. 

For the  determination of energy levels, radiative rates (A-values) and  oscillator strengths (f-values), El-Maaref  et al. \cite{elm} have adopted four different atomic structure codes, namely (i) the General-purpose Relativistic Atomic Structure Package (GRASP0: {\tt http://amdpp.phys.strath.ac.uk/UK\_APAP/codes.html}), (ii) its (another) revised version GRASP2K \cite{grasp2k}, (iii) the Flexible Atomic Code (FAC: {\tt https://www-amdis.iaea.org/FAC/}),  and finally (iv) AutoStructure (AS \cite{as}). This exercise, as recommended by several workers (\cite{fst},\cite{hkc},\cite{atoms}), is often performed for the assessment of accuracy and subsequently to have confidence in the generated data. Since each code differs in methodology, algorithm and/or the inclusion of relativistic effects, some differences in the generated data are often visible and understandable. Nevertheless, the comparison of energy levels shown in their table~1 is generally satisfactory, because the discrepancies are limited to $\sim$0.5~Ryd. Similarly, there are no large discrepancies for the A-values for many transitions, as shown in their table~2, although differences for one, namely 11--15 (4s4d~$^3$D$_3$ - 4p4d~$^3$F$^o_2$),  are up to three orders of magnitude. This is a very weak transition with f $\sim$ 2$\times$10$^{-7}$.

For the subsequent calculations of $\Omega$, El-Maaref  et al. \cite{elm} have mainly adopted the Dirac Atomic $R$-matrix Code (DARC), available at the same website as GRASP, i.e.  {\tt http://amdpp.phys.strath.ac.uk/UK\_APAP/codes.html}. However, they have considered two models, one with 13 levels of 4s$^2$, 4s4p, 4s4d and 4s4f, and another one, which includes additional 139 levels of the 4p4d, 4p4f, 4d4f, 4p$^2$, 4d$^2$, 4f$^2$, 4s5$\ell$, and 4p5$\ell$ configurations, i.e. 20 in total. In a collisional calculation these 152 levels were not manageable by them with their computational resources, and therefore they restricted  to the lowest 82, which generate 3321 transitions in total, but they have reported results for only 49 $\sim$1.5\%. For one of the transitions, namely 1--3 (4s$^2$~$^1$S$_0$ -- 4s4p~$^3$P$^o_1$), the background values of collision strengths ($\Omega_B$), differ by about a factor of four -- see their fig.~3. Since this is an allowed transition, its $\Omega$ values are easier to understand as these are directly related to the f-value and the energy difference, $\Delta$E$_{i,j}$. The $\Delta$E$_{i,j}$ for this transition are comparable, i.e. 6.7951 and  6.8950~Ryd in the smaller (13) and larger (152) models, but the f-values differ by $\sim$10\% and are  0.1566 and 0.1414 for the respective two models. Therefore, their results do not appear to be correct and we discuss these in more detail in section~3.

\begin{table}
\caption{Comparison of energy levels (in Ryd) for Zn-like W~XLV.} 
\begin{tabular}{rllrrrrrrrrrr} \hline
\\
Index &   Configuration & Level       &  GRASP1    & GRASP2     & FAC1     & FAC2    & GRASP0 & GRASP2K & FAC      \\ 
\hline \\												       
1   &	4s$^2$      & $^1$S$_0$        &  0.0000   &   0.0000	&  0.0000  &  0.0000 &  0.00  &  0.00	&  0.00    \\
2   &	4s4p	    & $^3$P$^o_0$      &  6.2297   &   6.3634	&  6.1802  &  6.3142 &  6.35  &  6.30	&  6.33    \\
3   &	4s4p	    & $^3$P$^o_1$      &  6.7951   &   6.8950	&  6.7433  &  6.8446 &  6.89  &  6.85	&  6.85    \\
4   &	4s4p	    & $^3$P$^o_2$      & 13.5717   &  13.6991	& 13.5378  & 13.6653 & 13.78  & 13.70	& 13.69    \\
5   &	4p$^2$	    & $^3$P$_0$        & 	   &  14.4624	& 	   & 14.5207 & 14.61  & 14.51	& 14.51    \\
6   &	4s4p	    & $^1$P$^o_1$      & 15.0174   &  15.0299	& 14.9754  & 14.9929 & 15.11  & 14.93	& 14.97    \\
7   &	4p$^2$	    & $^3$P$_1$        & 	   &  21.4856	& 	   & 21.3994 & 21.55  & 21.46	& 21.39    \\
8   &	4p$^2$	    & $^1$D$_2$        & 	   &  21.5764	& 	   & 21.4910 & 21.64  & 21.51	& 21.49    \\
9   &	4s4d	    & $^3$D$_1$        & 25.2925   &  25.4211	& 25.2360  & 25.3491 & 25.53  & 25.45	& 25.35    \\
10  &	4s4d	    & $^3$D$_2$        & 25.5426   &  25.6773	& 25.4850  & 25.6043 & 25.79  & 25.64	& 25.61    \\
11  &	4s4d	    & $^3$D$_3$        & 26.8388   &  26.9622	& 26.7851  & 26.8922 & 27.11  & 26.94	& 26.90    \\
12  &	4s4d	    & $^1$D$_2$        & 27.4817   &  27.3791	& 27.4228  & 27.3068 & 27.52  & 27.35	& 27.32    \\
13  &	4p$^2$	    & $^3$P$_2$        & 	   &  29.3747	& 	   & 29.2999 & 29.52  & 29.29	& 29.28    \\
14  &	4p$^2$      & $^1$S$_0$        & 	   &  29.7184	& 	   & 29.6448 & 29.87  & 29.64	& 29.63    \\
15  &	4p4d  	    & $^3$F$^o_2$      & 	   &  32.1293	& 	   & 32.0043 & 32.23  & 32.08	& 32.04    \\
16  &	4p4d  	    & $^3$D$^o_1$      & 	   &  33.4351	& 	   & 33.3055 & 33.58  & 33.22	& 33.30    \\
17  &	4p4d  	    & $^1$F$^o_3$      & 	   &  34.3697	& 	   & 34.2469 & 34.55  & 34.31	& 34.28    \\
18  &	4p4d  	    & $^3$P$^o_2$      & 	   &  34.3969	& 	   & 34.2716 & 34.52  & 34.31	& 34.27    \\
19  &	4s4f  	    & $^1$F$^o_3$      & 39.1760   &  38.9330	& 39.1461  & 38.8650 & 39.11  & 38.92	& 38.90    \\
20  &	4s4f  	    & $^3$F$^o_2$      & 39.0782   &  39.1840	& 39.0480  & 39.1208 & 39.37  & 39.19	& 39.15    \\
21  &	4s4f  	    & $^3$F$^o_4$      & 39.4596   &  39.4345	& 39.4321  & 39.3737 & 39.64  & 39.50	& 39.41    \\
22  &	4s4f  	    & $^3$F$^o_3$      & 39.8095   &  39.6804	& 39.7814  & 39.6181 & 39.89  & 39.65	& 39.64    \\
\\ \hline
\end{tabular} 

\begin{flushleft}
{\small
GRASP1: present calculations with  the {\sc grasp} code for 13 levels \\
GRASP2: present calculations with  the {\sc grasp} code for 152 levels \\
FAC1:  present calculations  with  the {\sc fac} code for 13 levels \\
FAC2: present calculations with  the {\sc fac} code for 152 levels \\
GRASP0: earlier calculations of El-Maaref et al. \cite{elm} with  the {\sc grasp} code for 152 levels \\
GRASP2K: earlier calculations of El-Maaref et al. \cite{elm} with  the {\sc grasp2k} code for 152 levels \\
FAC: earlier calculations of El-Maaref et al. \cite{elm} with  the {\sc fac} code for 152 levels \\
}
\end{flushleft}
\end{table}

\section {Energy levels and radiative rates}

Following El-Maaref  et al. \cite{elm}, we have performed two sets of calculations with 13 and 152 levels, described in section~1. Similarly, we have used both GRASP (GRASP1 and GRASP2) and FAC (FAC1 and FAC2) codes. The energies, and A-values, are comparable among these four calculations, and there are no appreciable discrepancies with their corresponding  work. Therefore, in Table~1 we compare all sets of energies for the lowest 22 levels alone, which include all 13 of the smaller model, and are also sufficient for our further discussion about other parameters. All energies listed in this table agree within $\sim$0.2~Ryd, see in particular levels 9--12, and results from smaller models (GRASP1 and FAC1) are comparatively less accurate because of the inclusion of (very) limited {\em configuration interaction} (CI). However, our present and earlier energies of El-Maaref  et al. (GRASP2 and GRASP0) also differ by up to 0.2~Ryd for some levels, and are perhaps due to the differences in the optimisation procedures, because both calculations use the same code and CI. Nevertheless, there is comparatively a better agreement (within $\sim$0.1~Ryd) among the GRASP2, GRASP2K, FAC2 and FAC calculations, which all use the same CI. Finally we will like to note that El-Maaref  et al. have incorrectly identified level 19 which should be $^1$F$^o_3$ and not  $^3$F$^o_3$ -- see levels 19 and 22 in their table~1, which are the same.

\section {Collision strengths}
 
Since our interest is in comparing $\Omega_B$  to verify the accuracy and reliability of  the results of El-Maaref  et al. \cite{elm}, we have performed calculations with FAC alone. This is also a fully relativistic code, as the DARC is, and mainly differs in the methodology because it uses the {\em distorted-wave} (DW) approximation in stead of $R$-matrix in the latter. Closed-channel (Feshbach) resonances can be naturally resolved in DARC calculations, as shown by El-Maaref  et al. in their figs. 1-3, but not in FAC, but these are not relevant for our present discussion. However, both approaches and codes are expected to produce comparable $\Omega_B$, for most transitions and over a wide range of energies, as has already been demonstrated in several of our earlier papers, including the one on tungsten, i.e. W~LXVI \cite{w66}. Nevertheless, this is only true when   similar number of levels  and their configuration state functions (CSF) are included in both calculations. 

In Fig.~1 (a, b and c) we compare our results for $\Omega$ from both models, i.e. 13 (FAC1) and 152 levels (FAC2) with those of El-Maaref et al. \cite{elm}, at energies {\em above} 30~Ryd, because resonances in this energy range are comparatively less dominant than at the lower one -- see figs. 1--3 of \cite{elm}. The $\Omega$ results of El-Maaref et al. are in a very limited energy range, up to only 64~Ryd, but are sufficient to give an idea about the (in)accuracy. Considered in this figure are three transitions, namely (a) 1--3 (4s$^2$~$^1$S$_0$ -- 4s4p~$^3$P$^o_1$), (b)  1--6 (4s$^2$~$^1$S$_0$ -- 4s4p~$^1$P$^o_1$) and (c) 2--4 (4s4p~$^3$P$^o_0$ -- 4s4p~$^3$P$^o_2$), and for clarity the $\Omega$ results of El-Maaref et al. have been {\em multiplied} by a factor of ten, in all three figures. The 1--3 and 1--6 are {\em allowed} transitions whereas 2--4 is forbidden. As stated earlier, and may also be noted from Table~1, the $\Delta$E$_{i,j}$ for both 1--3 and 1--6  transitions are comparable among different models and codes, and the f-values between the two models differ by only $\sim$11\% and 8\% for the respective two transitions, with a smaller model having larger f-value in each case. Therefore expectedly, the $\Omega$ from FAC1 are larger than those from FAC2, by nearly the same amount as their f-values are. Interestingly, the $\Omega$ of  El-Maaref et al.  are {\em lower} from the smaller model than the larger one -- see their fig.~3.  Additionally, the $\Omega$ of El-Maaref et al. are clearly {\em underestimated} by about two orders of magnitude. This is mainly because they have performed calculations with partial waves with angular momentum $J \le$ 9.5. This range  is {\em not} sufficient for the convergence, and as a consequence their results remain nearly constant, in stead of increasing with increasing energy, as seen in our work or expected for the allowed transitions. It may also be worth noting here that the inadequacy of such a small range of partial waves  has (already) been emphasised and demonstrated in several of our papers, such as  \cite{nixi}  for Ni~XI and \cite{w66} for W~LXVI. We discuss it further below.

\begin{figure*}
\includegraphics[angle=-90,width=0.9\textwidth]{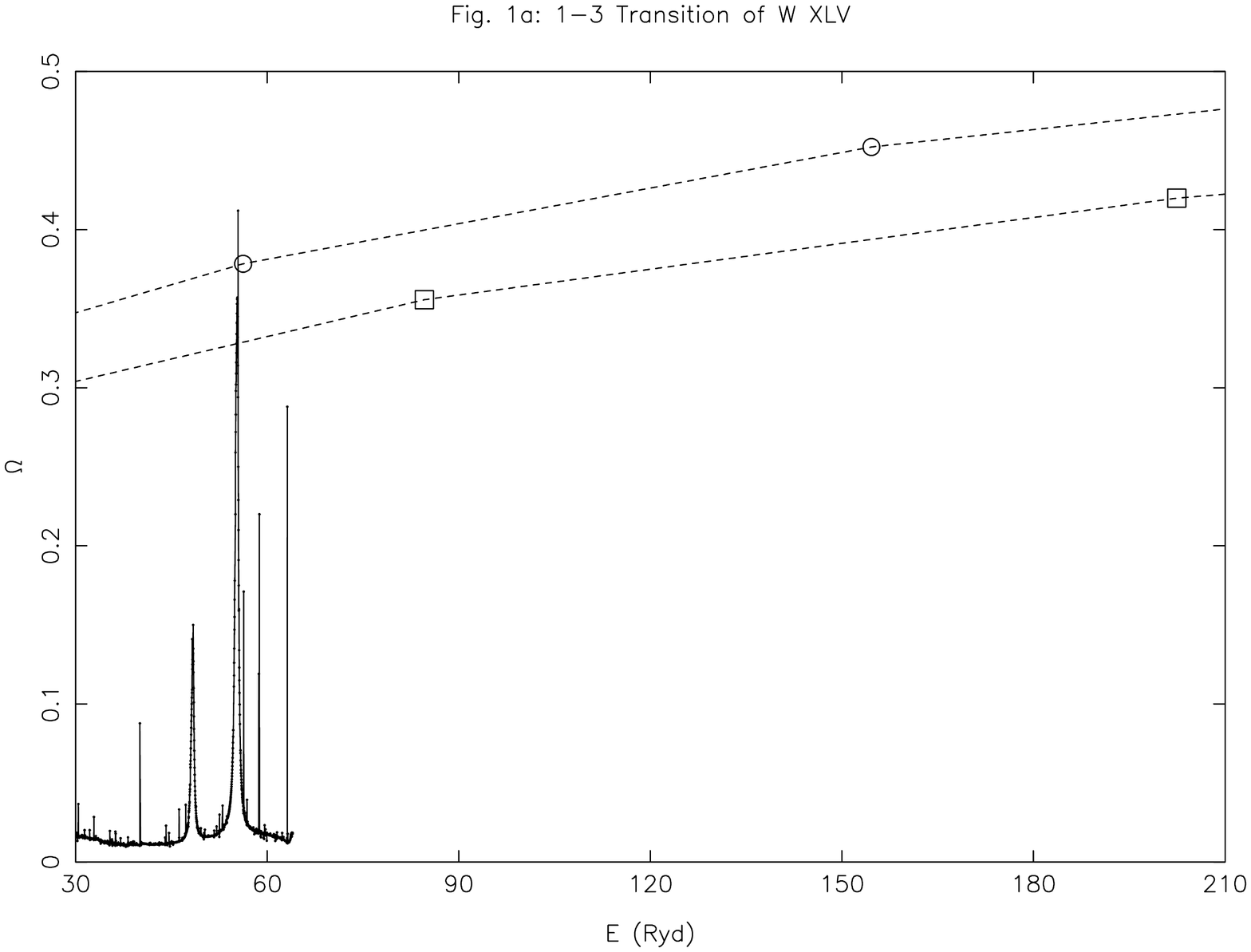}
 \vspace{-1.5cm}
 \end{figure*}
 
\setcounter{figure}{0}
 \begin{figure*}
\includegraphics[angle=-90,width=0.9\textwidth]{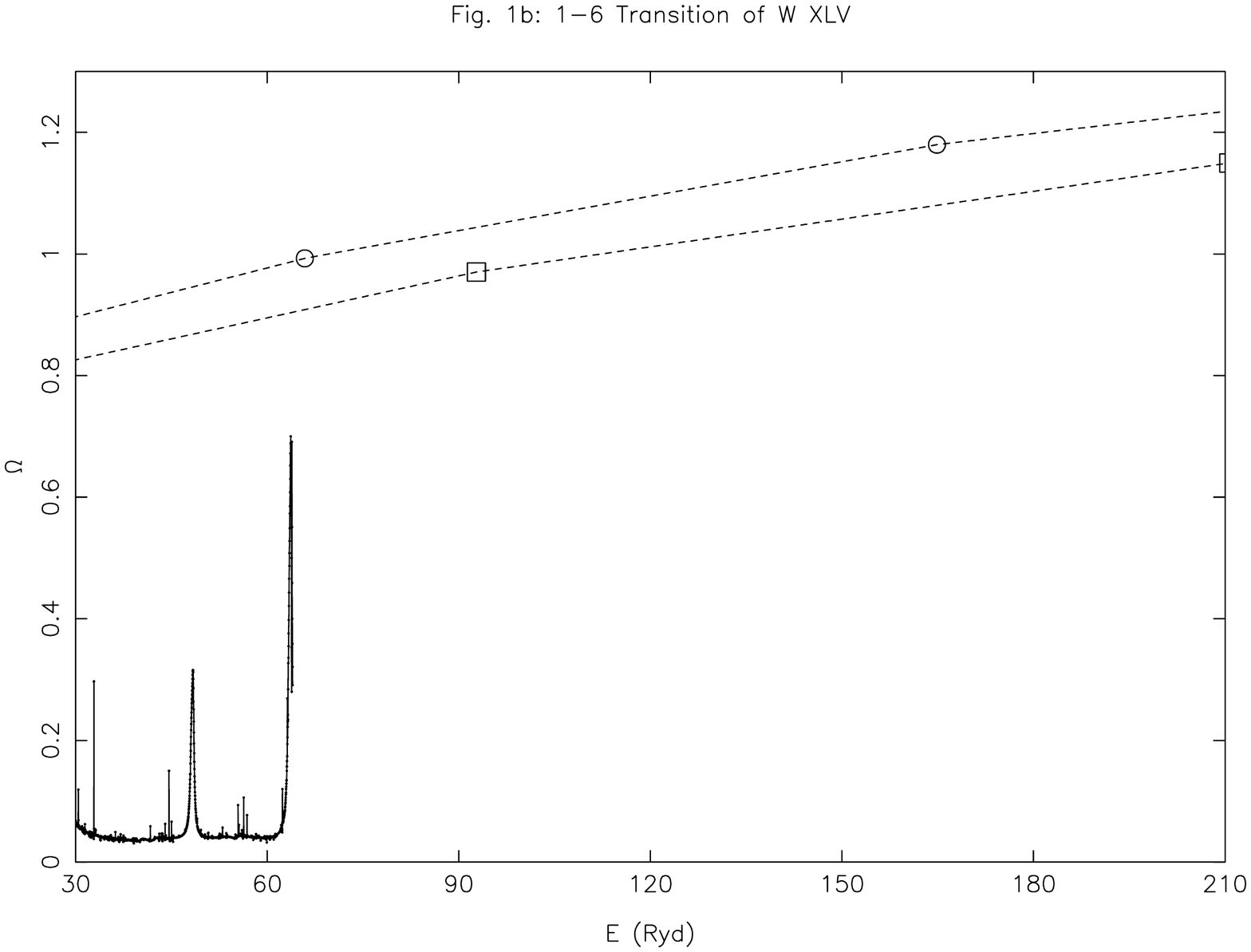}
 \vspace{-1.5cm}
 \end{figure*}
 
 \setcounter{figure}{0}
\begin{figure*}
\includegraphics[angle=-90,width=0.9\textwidth]{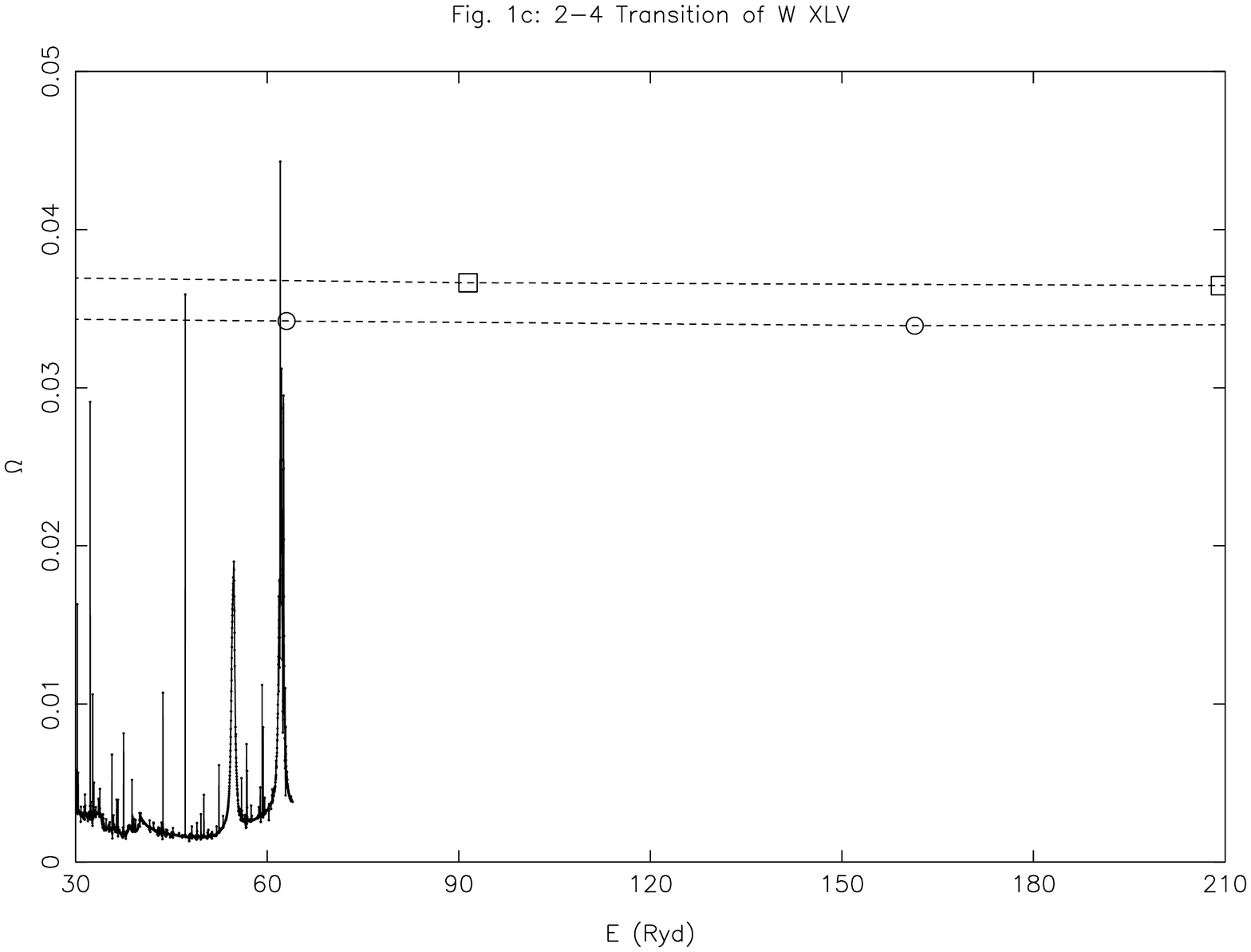}
 \vspace{-1.5cm}
\caption{Comparison of DARC and FAC values of $\Omega$ for  the (a)  1--3 (4s$^2$~$^1$S$_0$ -- 4s4p~$^3$P$^o_1$), (b)  1--6 (4s$^2$~$^1$S$_0$ -- 4s4p~$^1$P$^o_1$) and (c) 2--4 (4s4p~$^3$P$^o_0$ -- 4s4p~$^3$P$^o_2$) transitions of W~XLV. Continuous curves: earlier results of El-Maaref et al.\, \cite{elm} with DARC, broken curves: present results  with FAC, circles: 13 levels model and squares: 152 levels model. Results of \cite{elm} have been multiplied by a factor of ten.}
 \end{figure*}

It is well known that $\Omega$ for allowed transitions converge very slowly with $J$. However, even for a forbidden transition, such as 2--4 (4s4p~$^3$P$^o_0$ -- 4s4p~$^3$P$^o_2$), shown in Fig. 1c, $J \le$ 9.5 are highly insufficient for accurate determination, as the results of  El-Maaref et al. \cite{elm} are underestimated, again by two orders of magnitude, as seen for the allowed ones in Fig.~1a,b. Therefore, the $\Omega$ data reported by El-Maaref et al. are not only highly inaccurate but are also insufficient for any practical use. This is for many reasons, such as: (i) the energy range of their calculation is very (very) limited and does not even cover the entire thresholds region, (ii) their energy mesh to resolve resonances (0.02~Ryd) is very coarse and will result in overestimation of $\Upsilon$ (see \cite{ba48} and \cite{br27}), a parameter obtained after integration over an electron velocity distribution function, mostly {\em Maxwellian}, and required in the analysis or modelling of plasmas (see section~4), and (iii) the results being available for only 49 transitions among the possible 3321 belonging to the 82 levels considered by them, i.e. $<$1.5\%. Similarly, the differences in $\Omega_B$ shown by them in their fig.~3 for the 1--3 allowed transition between a smaller (13 levels) and larger (152 levels) models are not proportionate to the corresponding differences in the f-values, as expected and demonstrated by us in Fig.~1a. Besides this, there are other deficiencies in their work. For example, for inelastic transitions, such as 1--2 shown in their fig.~1, collision strengths are not possible at energies {\em below} threshold and they are {\em confused} between collision strength ($\Omega$) and collision cross-section ($\sigma$) -- see their figures 5--14 and the associated text.

 Finally, we will like to note that it is rather puzzling why  El-Maaref et al. \cite{elm} chose to calculate atomic data for an important ion like W~XLV when a much larger and more accurate data, and by the same codes as adopted by them (i.e. GRASP and DARC), have already been reported by Ballance and Griffin \cite{darc}, and are freely available on the ADAS website: {\tt http://open.adas.ac.uk/detail/adf04/znlike/znlike\_cpb07][w44ic.dat}. They have considered 168 levels, i.e. 152 of FAC2 (see section~1) plus additional 16 from the 4d5s and 4d5p configurations. We have also performed yet another calculation (FAC3) with these 168 levels but the results obtained are similar to those from FAC2, for all transitions considered above, and are therefore not being discussed any further. Furthermore, they have: (i) included contributions from partial waves with $J \le$ 35.5, (ii) have taken into  consideration the contribution from the higher neglected partial waves, through a top-up procedure, (iii) have calculated $\Omega$ up to very high energy of 1100~Ryd, required for the calculations of $\Upsilon$ up to very high  temperatures $\sim$ 10$^8$~K, which prevail  in fusion plasmas, and finally  (iv) have resolved resonances in a fine energy mesh of 0.002~Ryd, thus calculating $\Omega$ at over 50~000 energy points. Their complete set of $\Upsilon$ data for all 14~028 transitions among 168 levels, and over a wide temperature range, are available on the ADAS website, and are probably the most accurate available to date. However, for brevity they have not listed their results for $\Omega$ although resonances have been shown for a few selected transitions. Therefore, in the next section we make a brief comparison of their $\Upsilon$ results with ours from FAC.

\begin{figure*}
\includegraphics[angle=-90,width=0.9\textwidth]{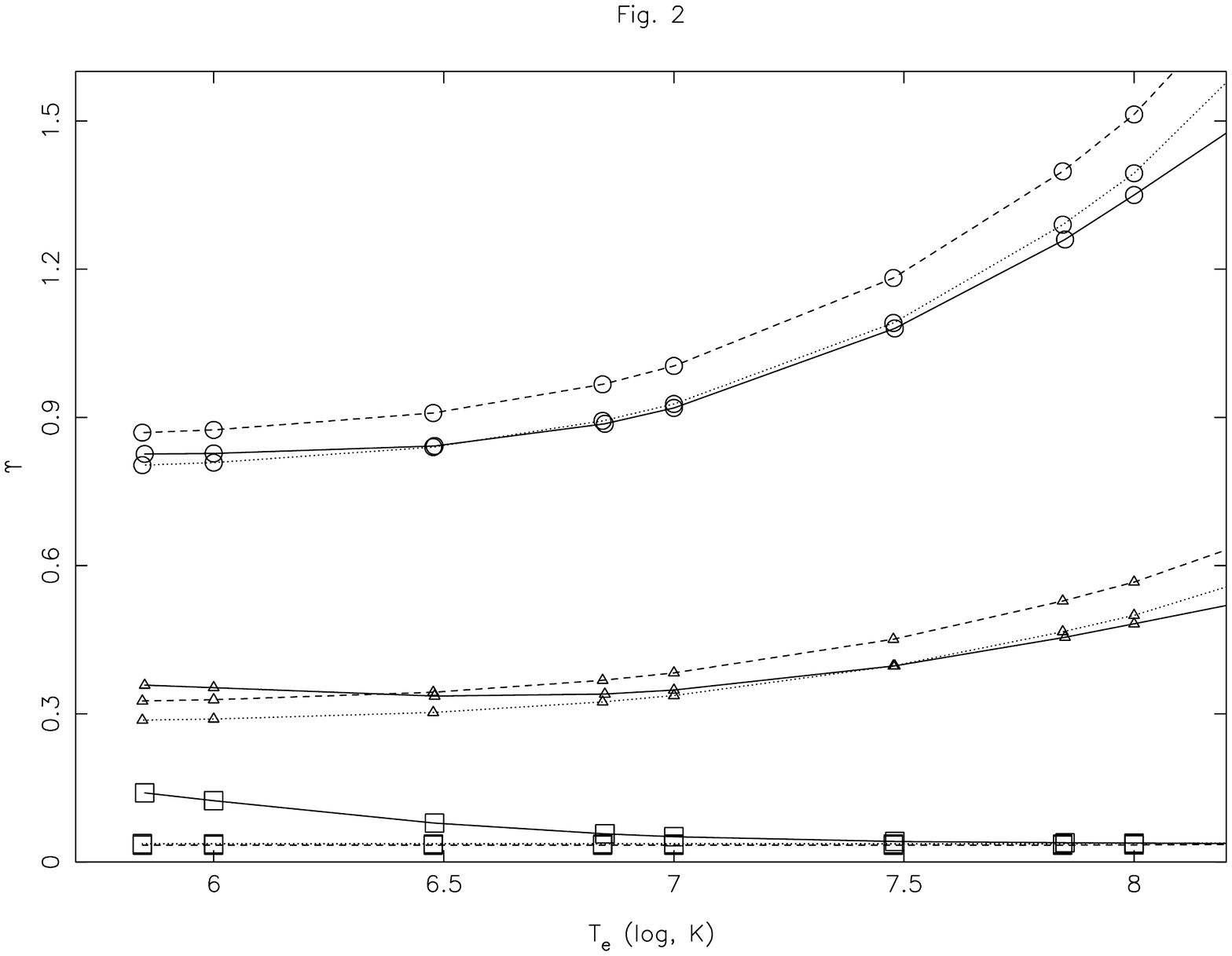}
 \vspace{-1.5cm}
\caption{Comparison of DARC and FAC values of $\Upsilon$ for  three transitions of W~XLV. Continuous curves: earlier results of Ballance and Griffin \cite{darc} with DARC, broken curves: present results  with FAC for 13 levels model, dotted curves: present results  with FAC for 152 levels model, triangles: 1--3 (4s$^2$~$^1$S$_0$ -- 4s4p~$^3$P$^o_1$),   circles: 1--6 (4s$^2$~$^1$S$_0$ -- 4s4p~$^1$P$^o_1$) and squares:  2--4 (4s4p~$^3$P$^o_0$ -- 4s4p~$^3$P$^o_2$) transition.}
 \end{figure*}
 
\section {Effective collision strengths}
 
As stated in section~3, the collisional parameter required for diagnostics or modelling of plasmas is {\em effective} collision strength ($\Upsilon$), and not $\Omega$. Therefore, in Fig.~2 we compare our results from FAC with those of Ballance and Griffin \cite{darc} from DARC, for the same three transitions, shown in Fig.~1. For comparisons results from both smaller (13) and larger (152 levels) models are shown. For the allowed transitions $\Upsilon$ results from FAC1 are slightly larger, as for $\Omega$, and for the reasons as already explained in section~3. For the 2--4 (4s4p~$^3$P$^o_0$ -- 4s4p~$^3$P$^o_2$)  forbidden transition there are no appreciable differences between the two models, and over the entire T$_e$ range.  Similarly, for allowed transitions there are no discrepancies between the two independent calculations (with similar levels being included), because both transition energies and oscillator strengths are comparable. However, for the forbidden transition the results of Ballance and Griffin are larger at temperatures below $\sim$10$^7$~K, and this is because of the resonances included by them but not by us. Therefore, this comparison confirms, once again, that the recent results of El-Maaref et al. \cite{elm}  are inaccurate, insufficient and practically useless.

\section{Conclusions}

In a very recent paper El-Maaref et al. \cite{elm} have reported (among other parameters) results for $\Omega$ for a few transitions of Zn-like W~XLV. Unfortunately, their results are 
underestimated by about two orders of magnitude, for both types,  forbidden and allowed. This is because they have considered a very limited range of partial waves with $J \le$ 9.5, and have completely ignored the contributions from higher neglected ones, which are not only significant and important, but are dominant in the determination of $\Omega$. This major deficiency of their work has been demonstrated through our independent calculations with FAC. For the benefit of the readers and potential users of data for this ion, we will like to note that a complete set of data for energy levels, A-values and $\Upsilon$ for transitions among 168 levels have been reported by Ballance and Griffin \cite{darc}, more than a decade ago, and their results are freely available at the ADAS website: {\tt http://open.adas.ac.uk/detail/adf04/znlike/znlike\_cpb07][w44ic.dat}. Since they have also performed fully relativistic calculations through the GRASP and DARC codes, and have resolved resonances in the thresholds region in a very fine energy mesh, their results are probably the best available to date, and can therefore be reliably and confidently applied in modelling or analysis of plasmas. Additionally, their results for $\Upsilon$ cover a very wide T$_e$ range and therefore are ideal for the studies of fusion plasmas. 

For many years a question often asked by the users of atomic data is about the accuracy, as generally several sets of data are available in the literature for any ion. It is not possible for every user to judge and assess the quality of the published data as s/he may not be competent and/or familiar with the fine details of a calculation. Neither is it possible for a single other qualified person (or a group) because a large amount of atomic data are almost continuously being generated and reported. Therefore, as has already been suggested \cite{fst}, it is upon the producers of data to make assessments of accuracy through rigorous checks and comparisons. It is a general tendency to believe that the latest reported data are more accurate than those already available in the literature, but it can be highly misleading as presently seen for the case of W~XLV, and several other ions earlier \cite{atoms}.  For the users of atomic data our considered advice is that a caution should always be exercised before using any data, otherwise the conclusions drawn may be faulty.


\end{document}